\documentclass{osa-article}

\journal{boe}

\articletype{Research Article}
\usepackage{float}
\usepackage{lineno}

\begin{document}

\title{High resolution TCSPC imaging of diffuse light with a one-dimensional SPAD array scanning system}

\author{E. P. McShane,\authormark{1,2,*} H. K. Chandrasekharan,\authormark{1} A. Kufcs\'ak,\authormark{2} \\ 
N. Finlayson,\authormark{2,3} A. T. Erdogan,\authormark{3} R. K. Henderson,\authormark{3} \\ K. Dhaliwal,\authormark{2} R. R. Thomson,\authormark{1,2} and M. G. Tanner\authormark{1,2}
}

\address{\authormark{1}Scottish Universities Physics Alliance, Institute of Photonics and Quantum Sciences, Heriot-Watt University, Edinburgh, UK\\
\authormark{2}Translational Healthcare Technologies Group, Centre for Inflammation Research, Queen’s Medical Research Institute, University of Edinburgh, Edinburgh, UK\\
\authormark{3}Institute for Micro and Nano Systems, University of Edinburgh, Edinburgh, UK}
\email{\authormark{*}epm1@hw.ac.uk} 



\begin{abstract}
We report a time-correlated single-photon counting (TCSPC) imaging system based on a line-scanning architecture. The system benefits from the high fill-factor, active area, and large dimension of an advanced CMOS single-photon avalanche diode (SPAD) array line-sensor. A two-dimensional image is constructed using a moving mirror to scan the line-sensor field-of-view (FOV) across the target, to enable the efficient acquisition of a two-dimensional 0.26 Mpixel TCSPC image. We demonstrate the capabilities of the system for TCSPC imaging and locating objects obscured in scattering media - specifically to locate a series of discrete point sources of light along an optical fibre submerged in a highly scattering solution. We demonstrate that by selectively imaging using early arriving photons which have undergone less scattering than later arriving photons, our TCSPC imaging system is able to locate the position of discrete point sources of light than a non-time-resolved imaging system.
\end{abstract}

\section{Introduction}
Two-dimensional single-photon detector arrays have been used to study time-resolved light passage through highly scattering media for medical imaging and spectroscopy \cite{Bruschini2019,CACCIA2019101}. In many cases, the light capturing ability of the imaging system is limited by detector characteristics such as fill-factor and total active area \cite{Intermite15}. This is often the case in two-dimensional (2D) single-photon avalanche diode (SPAD) arrays with integrated timing electronics, routinely deployed for time-resolved imaging due to their single-photon detection capability and high temporal resolution. This work presents the use of an alternate time-resolved imaging arrangement, applying a SPAD line array together with an imaging system which exploits one-dimensional (1D) scanning to form a 2D image. One-dimensional sensors can offer improved light capture ability over 2D sensors through higher fill-factor and larger active area designs \cite{Krstajic15}, as discussed in section 2.1. We implement a 512 $\times$ 1 pixel CMOS-SPAD line array \cite{RaII}, combined with a scanning optomechanical element to form a camera-like imaging system. This sensor and previous design iterations have been applied elsewhere for time-resolved spectroscopy \cite{KufcsakRAII}, optical coherence tomography \cite{Kufcsak:21} and pH sensing \cite{Ehrlich:17}. Photon arrival timing electronics are integrated within the CMOS sensor architecture for full time-correlated single-photon counting (TCSPC) measurements \cite{Becker}. Unlike many conventional 2D CMOS-SPAD imaging arrays, this line sensor boasts a high SPAD fill-factor (active detector area 28.2\% of total array area) as the integrated control electronics are placed off-pixel. Each pixel comprises 16 grouped SPADs to increase the photon capture area. The 512-pixel line-sensor defines the number of pixels in the acquired image along one axis, with number of pixels in the orthogonal axis dictated by a computer-controlled scanning element used to sweep the sensor across the FOV in the second axis. The advantage of this approach over using a 2D imaging detector array is that we effectively create a high resolution, large format imaging system, a 12 mm line (dimension of the sensor) combined with the arbitrary scanning positions used. While the need to scan in the second axis adds an overhead to image formation, this is more than compensated for by the ability to use larger optics for efficient light collection in a stand-off imaging modality relative to those appropriate when using typically smaller 2D detector arrays. \medskip

\noindent We outline the design and imaging ability of this scanning imaging system and demonstrate application to diffuse photon imaging of light emitted from discrete point sources submerged within a highly scattering phantom. When imaging with full TCSPC capability, the locations of a series of sequentially observed light sources are determined. The controlled illumination required to carry out this experiment is provided by a bespoke optical fibre probe machined using ultrafast laser ablation techniques, offering multi-point illumination at many discrete locations along the device length \cite{Chandrasekharan:21}. These two systems operating in tandem recover the location of 18 discrete point sources of light to track the otherwise obscured device length. We suggest the ability to locate fibre devices deep within scattering media is relevant to medical device location as previously described \cite{Mike}, we build upon this by offering improved light detection capability and device path observation rather than a single point. The improved time-resolved imaging of diffuse photons has application in further environments \cite{GOPAL1999331,Satat2016, LyonsMike, Radford:20}.

\section{CMOS-SPAD line scanning imaging system}
\subsection{CMOS-SPAD line sensor}

Line sensor architectures generally achieve higher fill-factors than 2D sensors, defined as the ratio of photo-sensitive area to overall focal plane array area   \cite{fillfactor}. A higher fill-factor can vastly improve the overall detection efficiency of a given system, and in this case, the fill factor improvement from a previously used 2D system (1.5\%\cite{Erdogan2015}) to the line scanning imaging system (28.2\%\cite{RaII}) is significant. Each line-sensor pixel contains 16 individual SPAD detectors orientated in an 8 $\times$ 2 sub-array format with a combined output. Individual detectors within a pixel can be optionally disabled in accordance with their characteristic background to maximise overall pixel signal to noise performance, allowing the user to remove the contribution of the highest noise SPADs in each pixel across the array. The sensor architecture contains two sets of 512-pixel arrays optimised for blue and red wavelength ranges \cite{RaII}, here we use the longer wavelength sensitive SPADs most appropriate for the near-infrared "optical window" of biological tissue near 800 nm \cite{nirwindow}, appropriate for a range of biophotonic applications. Examples of other recently developed SPAD line sensors can be found elsewhere \cite{Polimi,Oulu,VillaReview}.\medskip

 \noindent Single-photon detection timing is recorded with approximately 50 ps time stamping and 150 ps jitter on each pixel. Alternatively, the detector can operate in fast non-time-resolved photon counting or on-chip histogram creation modes suitable for other applications. Each line-sensor pixel is 116.365 $\mu m$ (height) $\times$ 23.78 $\mu m$ (width) leading to a total array length of 12.175 mm. The total size of the die in which the detector is cast is 12.648 mm $\times$ 1.990 mm in its long and short axis respectively. Owing to the larger length of this line sensor relative to the edge length of many 2D CMOS-SPAD arrays, larger optics can be integrated within the imaging system. Recent 2D imaging technologies have achieved impressive pixel numbers \cite{Morimoto:20,Ren:18}. Rather than comparing to all developments, we consider one such sensor ‘QuantiCam’ \cite{HendoQuanti} containing 192 $\times$ 128 single SPAD pixels, totalling 24,576 SPADs each with active area 22 $\mu m^2$. The total active area of the QuantiCam is  557,568 $\mu m^2$. Meanwhile, the sensor used in this work has 512 pixels of 16 SPADs totalling 8,192 SPADs for both the blue and red wavelength range optimised arrays with total active areas of 571,091 $\mu m^2$ and 399,512 $\mu m^2$ respectively. In each case we see that the total active area of the line-sensor is comparable to that offered by an array such as QuantiCam. An initial assumption is that these sensors would have similar photon counting capability if used with identical optical systems. However, significant in a widefield imaging application is the total chip dimension. The 12.175 mm dimension of the line sensor (compared to 3.2 $\times$ 2.4 mm chip dimension for QuantiCam) gives a "large format" sensor, for which larger optics can be used at a moderate working distance: here we use a 1-inch lens, with a numerical aperture of 0.5, at approximately 61 cm working distance to achieve a 30 cm FOV. Smaller sensor formats demand the use of a shorter focal length and thus (assuming similar NA) a smaller aperture lens at this working distance to achieve this FOV. As such, the light gathering power or ”etendue” using a large format sensor is much increased, allowing greater practical photon collection per second from a real scene. This is in addition to the discussed advantages of a large number of reconstructed image pixels, e.g. 512 $\times$ 512 = 262,144, an order of magnitude greater than QuantiCam. We have compared to QuantiCam as this chip has similar characteristics and was developed at a similar time to the line sensor used here. In this rapidly advancing field, higher pixel numbers and active areas are being achieved, although sometimes at the cost of lacking full TCSPC capability \cite{Morimoto:20}. We suggest that as line format sensors similarly develop, the potential advantages for imaging we describe should still be considered when comparing the newest technology.
 
 \subsection{Imaging system design}

A scanning mirror element sweeps the sensor FOV in the imaging axis orthogonal to the SPAD line array axis, achieving a 2D image composed of a series of sequentially acquired slices. To match the sensor pixel number requires 512 image slices in the second axis, resulting in a 0.26 Megapixel image. For a defined FOV, this offers improved image sampling relative to most 2D CMOS-SPAD imaging arrays. The scanning element consists of a two-inch reflecting mirror (Thorlabs, PF20-03-G01) mounted upon a rotation mount (Thorlabs, DDR25/M), offering high-speed movement (1800$^{\circ}$/s) and computer control \cite{MotorMatlab}. In principle this rotational element can scan a typical 30 cm dimension secondary axis (equating to an approximate total scan angle of 16$^{\circ}$) in 9 ms. Data acquisition is synchronised with mirror movement to automatically build image frames. A one-inch diameter lens (Thorlabs LA1951-B) mounted within an adjustable mount (Thorlabs SM1Z) achieves a 30 cm field of view along the line sensor axis at a working distance of 61 cm. The mirror is rotated a total of 16$^{\circ}$ to achieve a secondary axis with FOV equal to that obtained in the sensor hardware axis. The system also integrates a 2D CMOS imaging sensor (Thorlabs Kiralux, CS126CU) for conventional imaging of scenes when the mirror is rotated 90 degrees.

\begin{figure}[H]
\centerline{\includegraphics[width=1.1\columnwidth, trim=1 1 1 1,clip]{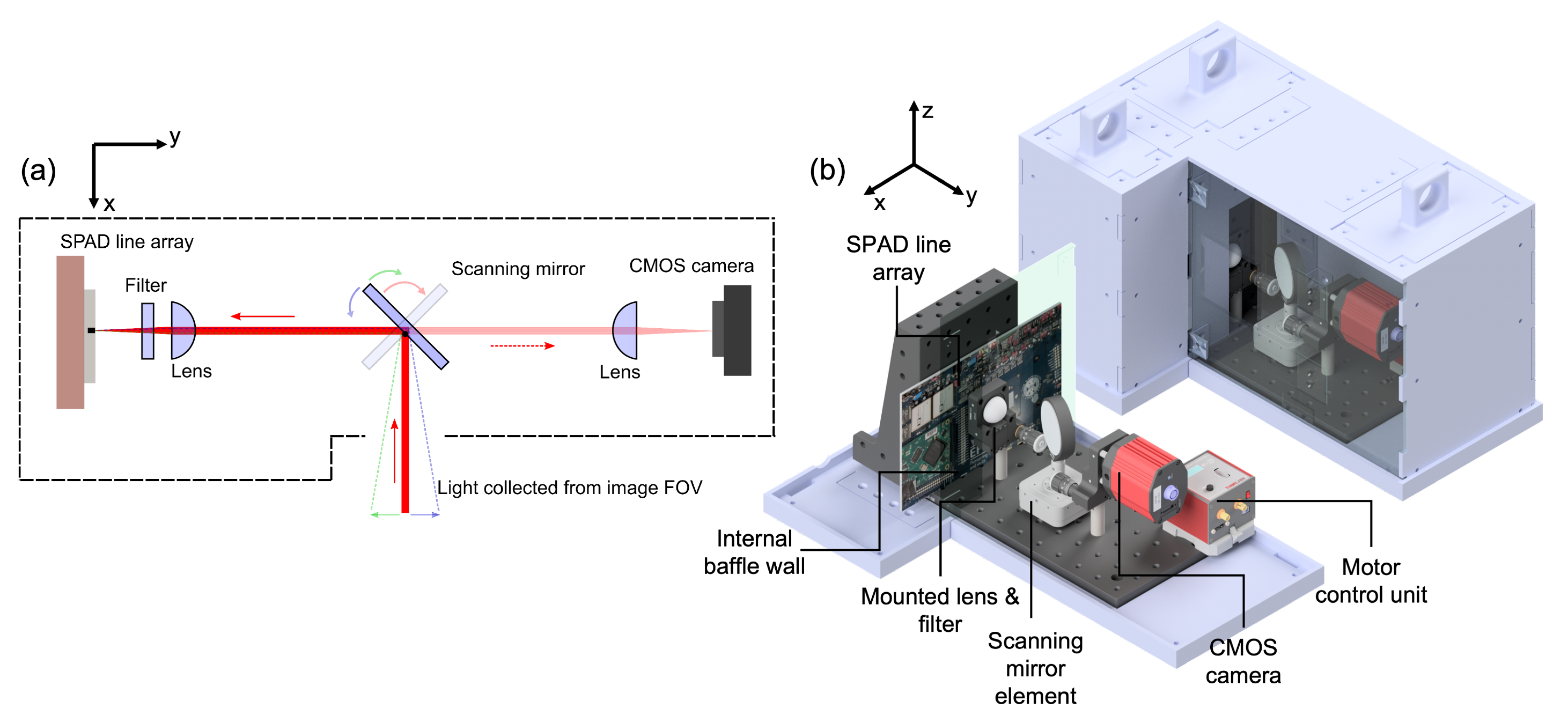}}

\caption{Optical design and housing of the scanning imaging system. (a) Outline of key system optical components. (b) Render of the imaging system housing and major internal components using computer aided design software. The housing was subsequently realised through 3D-printing, then foil lined, allowing the scanning imaging system prototype to be applied within ambient light scenarios. }
\label{fig:F1}
\end{figure}

\noindent Figure 1(a) outlines the key optical components of the system. Figure 1(b) displays computer-aided design renderings (Solidworks) of the imaging system outlining key system components. System hardware has been placed within a bespoke foil lined housing to mitigate against spurious light leakage, including an internal baffle with a narrow-band filter centred at 780 nm (Semrock, 780/12 nm BrightLine) directly in front of the sensor to limit incident ambient light being recorded during imaging. The housing containing these components allows the imaging system to be portable, rugged, and applied in off-the-bench scenarios. The total size of the system was minimised whilst also containing all system components, only requiring external connection for computational control and power, and to provide the TCSPC trigger signal. The system housing was developed using computer aided design package Solidworks and 3D-printed to specification.

\subsection{High-resolution imaging}

To demonstrate the TCSPC imaging capabilities of our system, we imaged a scene composed of several detailed objects (Figure 2(a)) which were flood illuminated from a fibre laser source (785 nm, 0.5 mW, 96 ps pulse duration) which was placed adjacent to the imaging system.  Photons reflected/scattered from the scene were collected for a total exposure time of 10.24 seconds, encapsulating the entire measurement of all image strips. The total exposure time was limited by requiring sufficient photon counts to achieve satisfactory signal-to-noise, and not by scanning electronics. Each image strip was produced with a frame exposure time of 20  $\mu s$ (in which a single photon can be counted and time tagged on each pixel), and 1000 frames sequentially acquired at each mirror position. Time-resolved images separated by a temporal step of 0.1 ns can be seen within Figure 2(b-i), a subset of the frames acquired at 50 ps separation. Details such as the well-defined text can be clearly distinguished, and the 3D profile of the target objects is represented frame to frame. This clarity of imaging is achievable with high image sampling due to the effective high pixel density created by sweeping the line sensor across the target. Imaging resolution is 0.59 mm based on pixel density (30 cm / 512 pixels), imaging a test target confirmed imaging resolution is limited by pixel pitch rather than optical system resolution. Within Figure 2 the presence of some fixed pattern noise is apparent, these effects are due to pixel-to-pixel variation in dark count rate and sensitivity. Sensitivity was not calibrated and corrected for within this imaging, however correction could be applied in future applications.

\begin{figure}[H]
\centerline{\includegraphics[width=1.2\columnwidth, trim=5 1 5 5,clip]{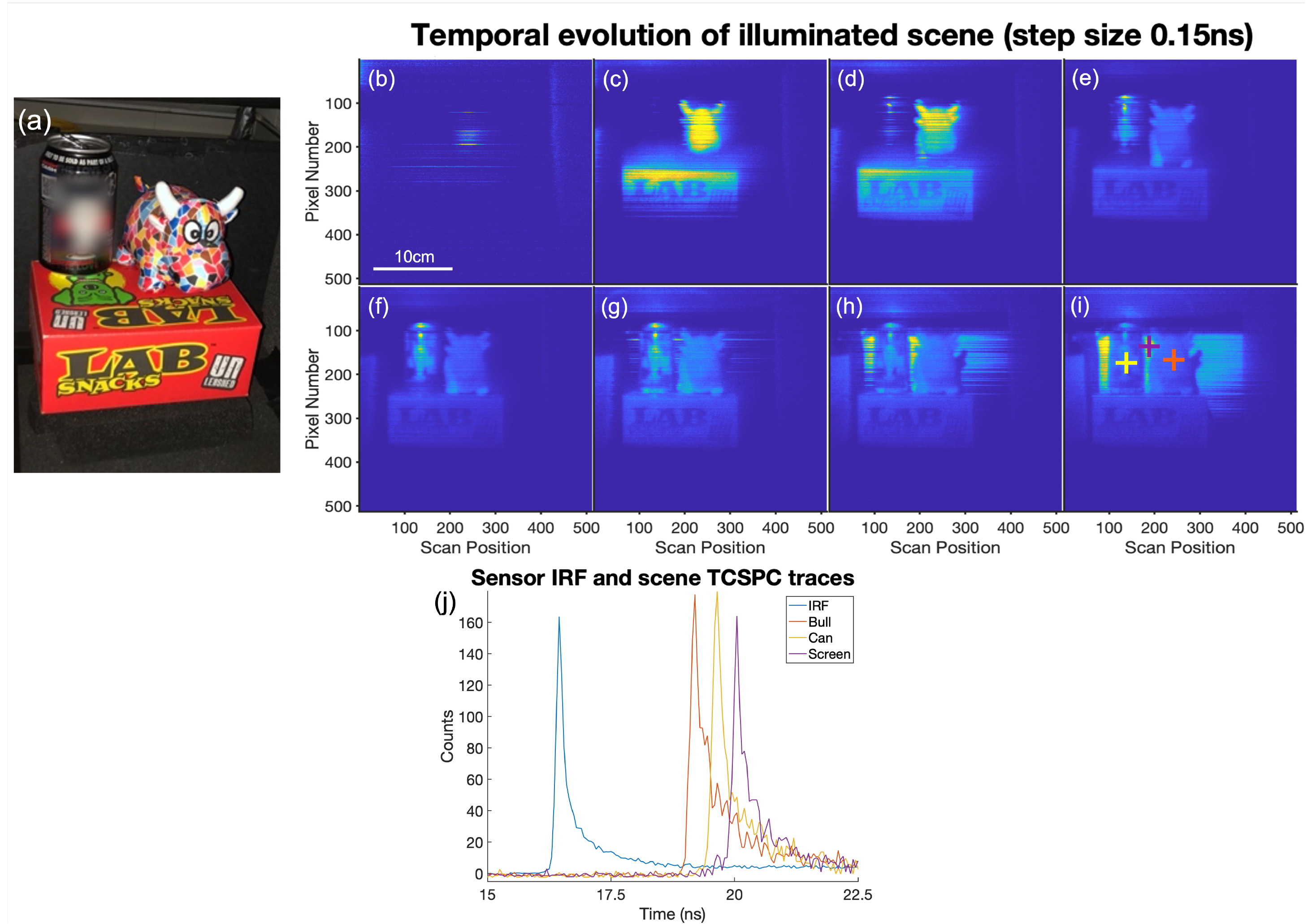}}
\caption{TCSPC imaging of a flood illuminated scene. Objects are disambiguated in time, revealed only as the illumination photons encounter them within the measurement window. (a) Camera image of the scene. (b-p) Images at increasing times within the measurement window (step size between images of 0.1 ns). (j) TCSPC traces for direct sensor illumination to obtain the instrument response function (IRF), and the reflected signal extracted from the three locations marked in (i). (See Visualisation 1).}
\label{fig:F2}
\end{figure}

\noindent In the scene above each tile shows the light detected at a specific time-point within the measurement window, the series describes the light washing over the scene. Objects placed closer to the light source are illuminated earlier with details behind becoming visible later. For example, two of the objects are illuminated at slightly earlier times, e.g. (b)-(d), in contrast to the light pulse reaching the back of the scene later in the time window, e.g. (h)-(i). Extracted timing histograms from three locations within the scene are shown in (j), along with the instrument response function (IRF) for direct illumination of the sensor. Fine temporal resolution can be used to selectively observe different objects within the field, limited by the IRF. Well established analysis could be used to perform depth imaging or light detection and ranging (LIDAR) \cite{Tobin2021,Mora-Martin:21,McCarthy:13}, including imaging partially obscured objects\cite{Maccarone:19}. 

\section{Time-resolved diffuse photon imaging }
 
 Time-correlated imaging can be used to observe the progression of light through highly scattering media for imaging applications  \cite{GOPAL1999331,Satat2016,LyonsMike, Radford:20}. For example, time-resolved imaging of light emission from a point source suspended within scattering media has the potential to infer the location of that source  \cite{Mike}. As photons pass through a diffusive phantom media, they undergo many scattering interactions with constituent particles. This acts to spread the pulse of photons in both space and time. An imaging system focused upon the edge of a phantom, where photons exit to free-space, will observe a region of scattered light emission. Temporal observation of photons emerging from the medium will show a significant variation with respect to an unperturbed laser pulse travelling through a media with minimal scattering. Many scattering events will result in a time broadened laser pulse, the peak of which will correspond to the modal optical path length of photon transit through the media, with associated transit time at which the greatest number of photons will arrive at the sensor. Photons emerging relatively late and early (with respect to the temporal peak) within the measurement window are considered to have statistically undergone more and less scattering respectively. More scattering events generally act to increase the optical path length travelled, and therefore delay photon transmission. \medskip
 
 \noindent At early times in the measurement window the subset of photons emerge that have undergone minimal scattering (relative to the later arriving photon cloud) having taken the shortest optical path from the source through the phantom  \cite{Tuchin_1997}. However, fast transit along a short optical path is also possible with many scattering events, provided scattering is in a forward direction, approximately maintaining the same direct path. This is similar in practice to photons with few scattering events, also following such a direct path. Many varieties of biological media exhibit large amounts of anisotropic scattering, such that scattered photons traversing through the media have a high probability of being forward-scattered \cite{Jacques_2013}, consistent with the theory of scattering from spherical particles. As such, more photons have an approximately direct path than just those minimally scattered. Photons emerging first from the scattering medium (early arriving photons at the imaging system detector, “early-photons”) can be temporally separated during data processing from the later arriving photon cloud. Such early-photons represent the exit point of photons that have taken the most direct path through the scattering medium. This location can be used to infer the obscured light source position \cite{Mike}, as studied below. \medskip
 
 \noindent A scattering phantom composed of homogenised semi-skimmed milk (a colloidal lipid suspension) provided a scattering coefficient comparable to that of biological tissue. The broadband optical properties of milk are dominated by scattering, characterisation of the optical properties of milk along with information on photon interaction with lipid particles can be found elsewhere  \cite{Milk}. A phantom solution of 1:4 parts semi-skimmed milk (1.7\% fat) to water was expected to provide a reduced scattering coefficient comparable to the 9 cm$^{-1}$ of muscle tissue reported within  \cite{Ntziachristos2010}. A bespoke polymer optical fibre was anchored in a fixed location to provide highly non-directional point source illumination (785 nm), the emission profiles are described in detail elsewhere \cite{Chandrasekharan:21}. The phantom tank dimensions were 40 $\times$ 25 $\times$ 25 cm (25 L) with illumination suspended approximately 15 cm from the imaged side of the tank, shown in Figure 3(a,b). Laser power launched into the fibre was 4.5 mW, based upon previous measurements of light emission by Chandrasekaran et al.\cite{Chandrasekharan:21} we estimate emission from a discrete point on the fibre of 2.5-3.3 mW. This is expected to be emitted and to propagate homogeneously in the phantom.
  \begin{figure}[H]
\centerline{\includegraphics[width=1\columnwidth, trim=12 4 5 5,clip]{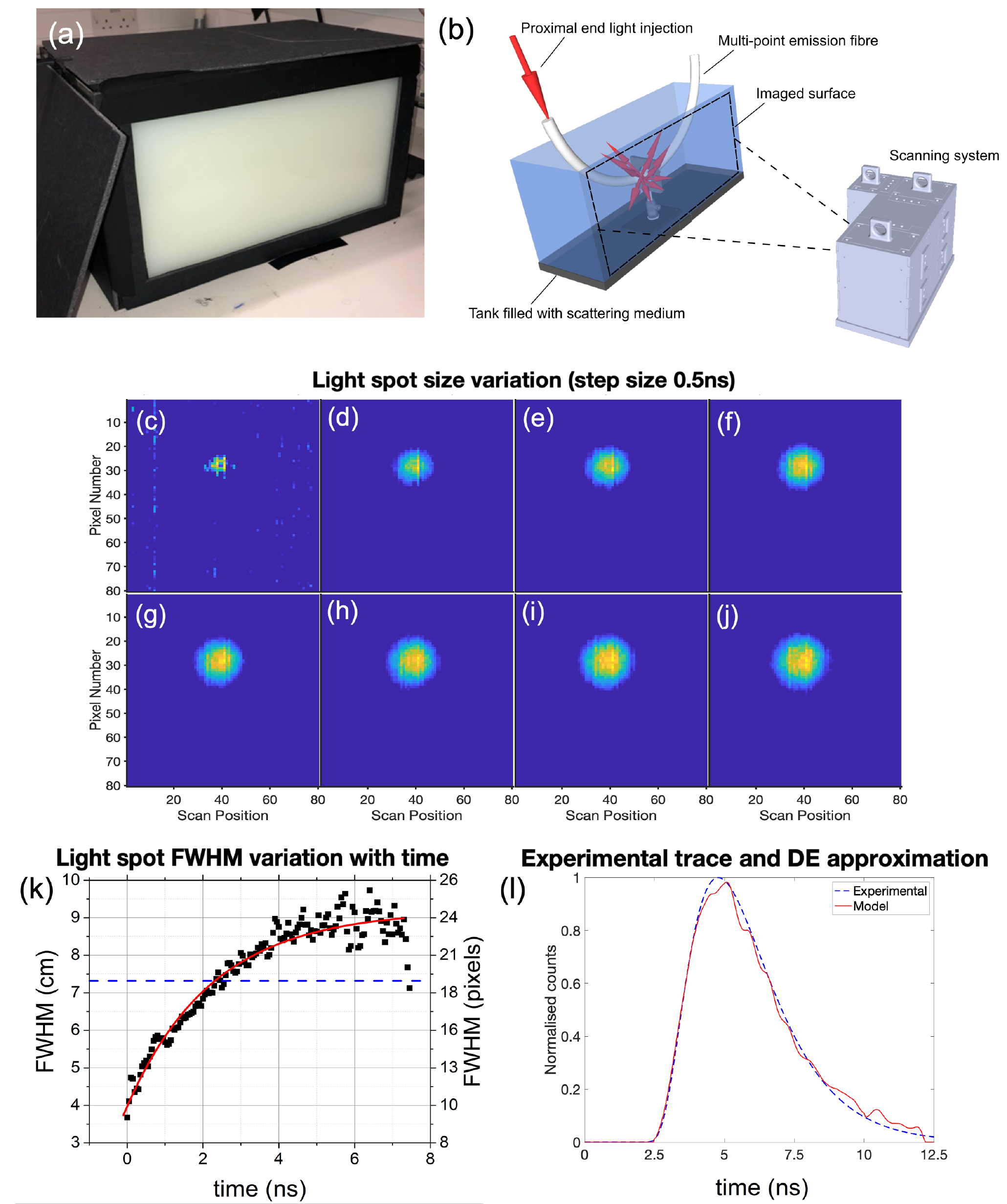}}

\caption{Temporal evolution of light emitted from a single obscured point source located within the phantom. (a) Camera image of the scattering phantom described further within the text. (b) Schematic of the experimental arrangement. (c-j) Variation of the imaged photon spot size with time. Each tile is separated in time by 0.5 ns, ten 50 ps time bin steps. (k) Temporal variation in the full width half maximum of a vertical strip taken through the imaged spot (a subset of which are shown in Figure 3(c-j). The FWHM of the spot is seen to increase dramatically. The red trend line is plotted as a guide to the eye, proportional to  ($1-e^{-kt}$). The FWHM of an all-photon image containing all light detected summed together is plotted as the blue horizontal line. (l) TCSPC histogram for the centre of the spot, compared to theory, yielding an approximate reduced scattering coefficient of 9 $cm^{-1}$ and absorption coefficient of 0.11 $cm^{-1}$. }
\label{fig:time grid}
\end{figure}
 
\noindent The exposed side of the tank was imaged with the system described, with a per frame exposure time of 100 $\mu$s for 2000 frames in each of the 160 image slices, leading to a total image acquisition time of 32 seconds. For the purpose of this experiment, recovered images were down sampled to improve signal to noise, hence the reduction in pixel density observed in Figures 3, 4 and 5 with respect to the pixel density presented in Figure 2. The observed signal evolution from a point suspended within the scattering phantom is presented within Figure 3(c-j). The light spot size varies significantly within the measurement window as photon scattering progresses with the full width at half maximum (FWHM) of the imaged light spot increasing as presented in Figure 3(k). The FWHM of the non-time-resolved sum of all light within the measurement time window has been plotted as a horizontal line. The photon emission spot is, as expected, more tightly located at early time points than both the non-time-resolved “all-photon” sum and the later arriving photon cloud. Time resolved observation of photon transit allows recovery of the phantom’s optical parameters \cite{Mosca:20}. Fitting data to an established model \cite{Patterson:89} as shown in Figure 3(l) confirmed scattering and absorption coefficients similar to those of various biological tissues \cite{Mosca:20}.  \medskip

\noindent From these simple experiments it is clear that the exit locations of the later arriving photons are less representative of the position on the tank surface that is closest to the source. Therefore, we apply an early-photon processing rationale to estimate the light source location with increased confidence. However, this homogeneous model is a poor mimic of biological tissue, and the later arriving photon cloud also provides a clear suggestion of the original source location. This would not be the case with increasing amounts of optical complexity in biological models.

\subsection{Device length location in a scattering phantom}

The scanning imaging system was applied to observe diffuse light emitted from many discrete point sources submerged within the scattering phantom. The combined spatial and temporal resolving capability was utilised to infer the spatial location of a series of light emission points along a polymer optical fibre device. The bespoke multi-point emission optical fibre provided known illumination from eighteen discrete selectable points as outlined within  \cite{Chandrasekharan:21}, placed along the fibre device length with a pitch of 2 cm. The points were sequentially illuminated and imaged while submerged in the scattering phantom, with the fibre curving in an arc across the FOV. Early and late emerging photon images are compared in Figure 4 for nine of the emission points, while the fibre path is analysed in Figure 5, including the path of the fibre observed without a scattering medium. \medskip

\noindent We apply TCSPC imaging to locate each light source submerged within the phantom by analysing the time-resolved images of each point (comparable to that seen within Figure 3(\textcolor{red}{c-j})), taking the earliest light as most representative of the location. We track the fibre device path across the phantom, as the discrete point sources span the device length. Image acquisition, control of illumination point, and data processing are automated to predict the spatial location of each light source on a point-to-point basis. Data processing (not in real time) proceeds as follows. Firstly, the peak of photon counts of the TCSPC histogram (detailing the distribution of photon arrival times) was found from each illumination point data set. The time-point at which comparatively earlier photons were detected within the measurement window was then defined as when the signal reached 10\% of the peak photon counts. The image of the photon spot at this early arrival time was extracted, and the centroid (arithmetic mean position) of the bright region was taken as an estimate of the light source location. It should be noted, earlier time points given by a lower photon threshold produce a smaller photon cloud image for determining source location. However, here a compromise was made to provide sufficient signal to noise to allow reliable automatic processing of data. In future more advanced processing of earlier photon arrivals, or indeed analysis including later photon arrivals \cite{Radford:20}, could be employed.

  \begin{figure}[H]

\centerline{\includegraphics[width=0.95\columnwidth, trim=1 1 1 0,clip]{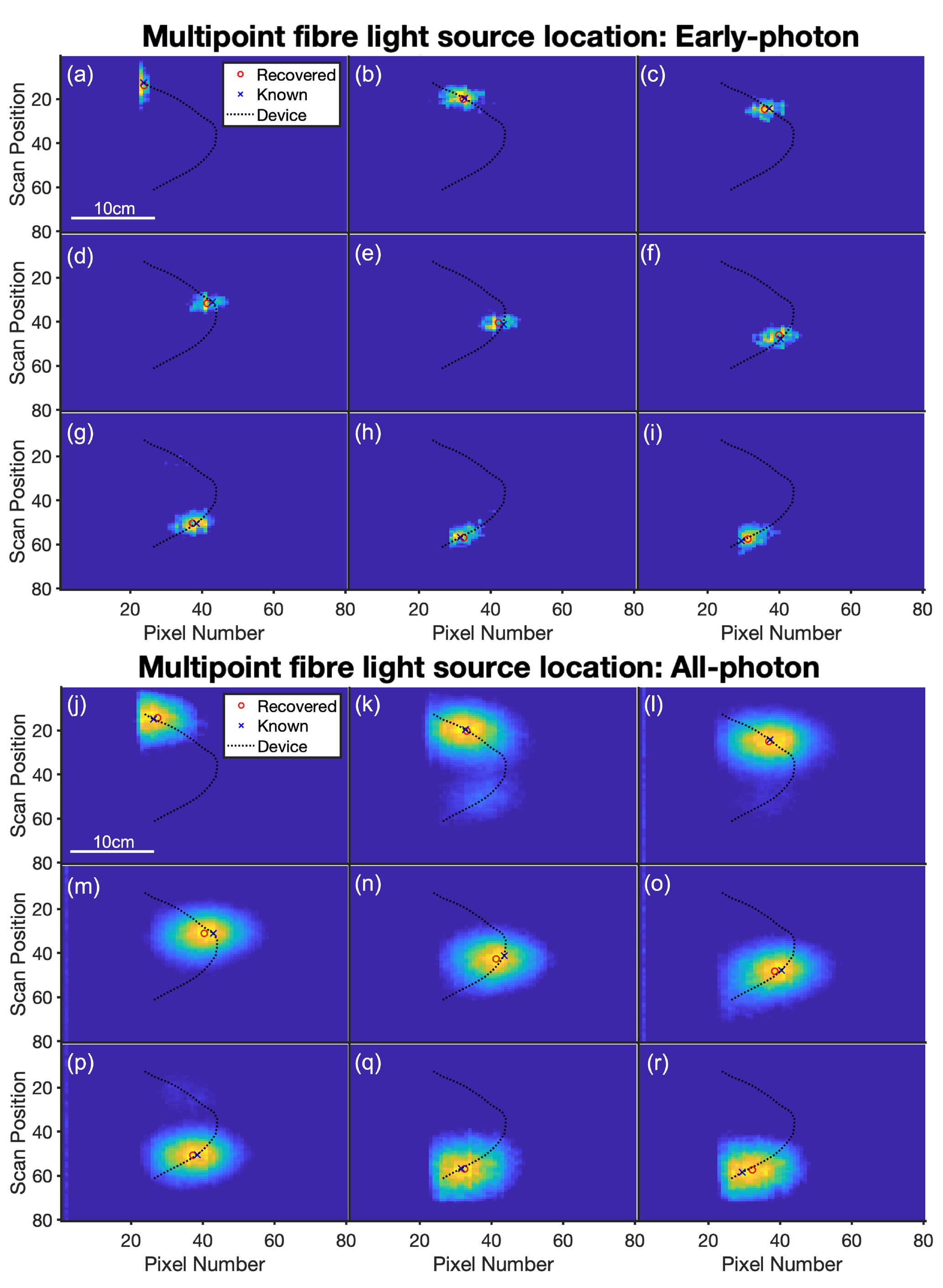}}

\caption{Early- and all-photon images of light emitted from nine discrete fibre point sources submerged within the phantom. The known device path (imaged in the absence of scattering particles) has been overlaid on top of each data-set in black. (a-i) Early-photon and (j-r) all-photon images}
\label{fig:F4}
\end{figure}

\noindent Figure 4 displays the early-photon and non-time correlated all-photon images of light emerging from the tank emitted by nine discrete point sources along the fibre device length. The early-photon emission spots are observed to closely track the arc of the optic fibre, with the observed emission tightly confined to a small area, whereas the all-photon images show larger emission regions. The recovered locations were compared to the known light-source locations, imaged in the absence of scattering elements within the phantom (Figure 5(a)). The recovered and known locations are shown within Figure 5(b), where all point locations (blue stars) were recovered in proximity to the known locations (red triangles). The average deviation between the two locations across the dataset was small, at approximately 0.51 cm (1.37 image pixels of Figures 5(b,c)).

   \begin{figure}[!ht]

\centerline{\includegraphics[width=1.25\columnwidth, trim=1 2 1 2,clip]{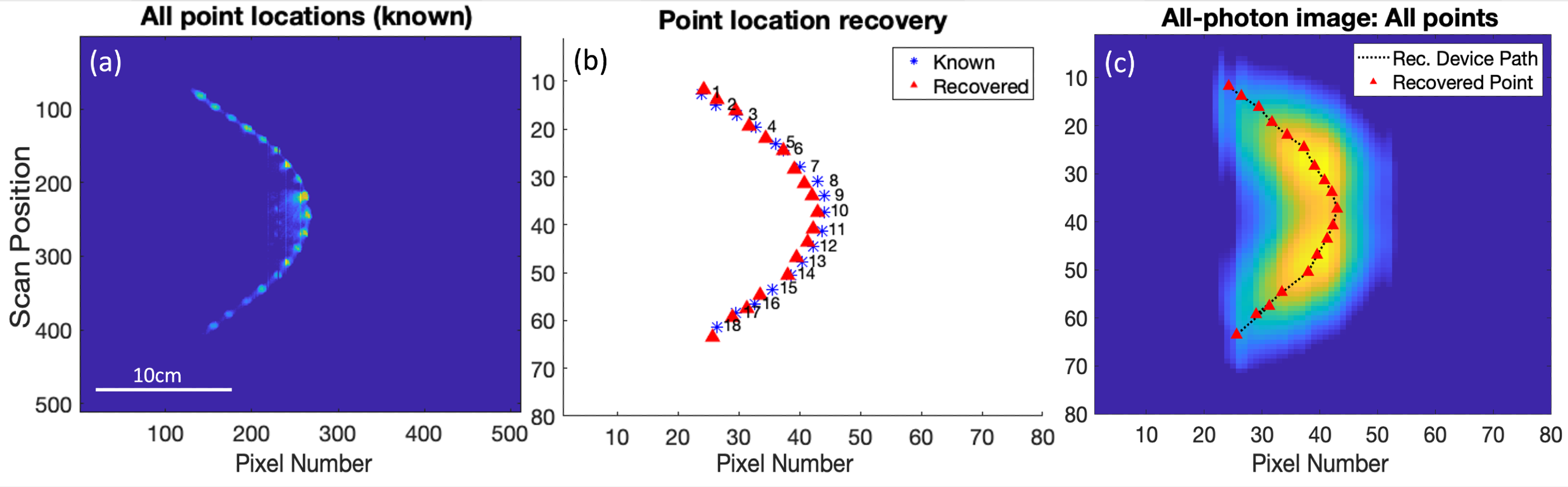}}

\caption{Obscured device tracking using TCSPC imaging. The known emission point locations (imaged in the absence of scattering particles) have been recovered using the early-photon processing rationale outlined above. (a) Combined visualisation of eighteen point sources across the phantom, imaged in the absence of scattering particles. (b) The locations recovered using automated processing (red triangles) along with the known locations from (a) (blue stars). (c) Combined image of all light emitted from all sources on the optical fibre, compared to the recovered locations. (Note: The resolution in image (a) is higher than the binned images used for location recovery processing). (See Visualisation 2).}
\label{fig:F5}
\end{figure}
 
 \noindent Figure 5(c) has been created using a summation of the eighteen all-photon images for each point source (e.g. the images in Figure 4(j-r)). When all images are overlaid in this fashion a poor correlation to the recovered locations (plotted in red) is observed, representing attempting to locate the device with neither the bespoke controlled fibre illumination nor time-resolved imaging capability. The authors note there may be some systematic variation in the recovered and known locations, which may be a result of boundary effects at the edge of the tank. \medskip
 
 \noindent The scanning imaging system and multi-point fibre device used in combination demonstrate device tracking within the highly scattering phantom with sub-centimetre accuracy. The high spatial resolution of the imaging system is used to provide discrete information of light-source location over centimetre incremental steps throughout the FOV, and the high temporal sensitivity of time-stamping is critical in carrying out the early-photon analysis outlined above. Using this methodology an entire fibre device, spanning approximately 36 cm was tracked within a highly scattering phantom.
 
 \section{Conclusion and future considerations}

 We present an imaging system which enables a CMOS-SPAD line array to be used for 2D time-resolved imaging. The rugged and portable design of the system makes it applicable for imaging both on the bench and from suspended locations. The system has been shown to offer full TCSPC imaging with comparatively high spatial resolution and precise temporal resolution over a wide-field. Image acquisition speed is not limited by mirror scan rate, which could in principle move at ~100 Hz over the full field of view if neglecting mirror acceleration overheads. A galvo scanning mirror could instead be implemented and may be more suited to rapid scanning. However applications are likely to be rate limited by the photon flux. As discussed this format of line sensor is advantageous for efficient photon collection as compared to scanning a single point sensor on multiple axis – the photon counting burden is instead spread over 512 detectors in parallel. Further the large line dimension is favourable for optical collection as compared to similar detectors arranged in two dimensional arrays, at the cost of additional system complexity. \medskip
 
  \noindent Therefore this work describes an effective, practical, adaptable, and relatively simple approach to wide field time resolved imaging of benefit to the wider community. Here optical arrangement is chosen to enable a wide field of view (30 cm) at moderate working distance (61 cm), advantageous for a number of applications but adaptable as needed. Moving to higher imaging rates would be important for applications such as LIDAR in self driving cars. This would also benefit from line scanning illumination efficiently performed with the same scanning mirror, thus maximising returned signal versus emitted laser power. In practice a continually rotating (rather than scanning) mirror might be used, but the principle is the same as this work. \medskip

 \noindent The system here has been optimised for biophotonic applications working with laser sources with wavelengths within the optical window for biological tissue, in this case, 785 nm. A detailed scene was used to exhibit the high spatial resolution of the system, useful for imaging within complex environments. This could be of use, for instance, during fluorescence guided surgery \cite{Stewart_2021} where medical indicator substances such as indocyanine green (ICG) are introduced intravenously to build up in target tissues such as lymph nodes. ICG fluorescence in the near-infrared can then guide surgery, time resolved fluorescence allows better separation of the target signatures from endogenous tissue fluorescence. Similar techniques have been demonstrated with a point probe to characterise tissue autofluorescence lifetimes and observe location of malignancies during surgery \cite{Marsden:20}. \medskip
 
 \noindent The system here was also used to locate the path of an optical fibre device across a highly scattering phantom, retrieving the spatial location of a series of discrete point sources. Eighteen light source locations along the fibre length were estimated with sub-centimetre accuracy with respect to their known location deduced in the absence of scattering particles. The results from this study provide evidence for the use of these two devices in tandem to track an entire fibre device within scattering media. Future iterations of this work will apply the technology in a more complex regime to better mimic real-biological conditions containing interfaces between media types where the extent of multiple scattering will more significantly alter the received time-correlated images, potentially providing a sterner test of the location rationale outlined above. However if successful this describes a method to guide medical procedures placing catheters modified to contain the optical fibre. This optical technique would reduce the reliance on X-rays, improving clinical workflow and healthcare outcomes.

\begin{backmatter}
\bmsection{Funding}
\noindent This work was funded by the UK Science and Technology Facilities Council (STFC) (ST/S000658/1, ST/S000763/1). EMS thanks the STFC for support through an STFC Ph.D studentship. The authors thank the EPSRC for support through EP/R005257/1.

\bmsection{Acknowledgments}
\noindent The authors would like to thank Conan Bradley Product Design for their expertise which was critical in the realisation of the imaging system housing design. The authors also thank Thorlabs inc. for granting permission to use their branding within this work.

\bmsection{Disclosures}
\noindent The authors declare no conflicts of interest.

\bmsection{Data Availability Statement} \noindent Data underlying the results presented in this paper are available on the Heriot-Watt server \cite{DataHW}.

\end{backmatter}

\bibliography{sample}






\end{document}